\documentclass[12pt, a4paper]{article}
\usepackage[utf8]{inputenc}
\usepackage{hyperref}
\usepackage{geometry}
\usepackage{graphicx}
\usepackage{subcaption}
\usepackage[numbers]{natbib}
\usepackage{amssymb}                                                                                                                                                                                                                                                                                                                                                                                                                                                                                                                                                                                                                                                                                                                                                                                                                                                                                                                                                                                                                                                                                                                                                                                                                                                                                                                                                                                                                                                                                                                                                                                                                                                                                                                                                                                                                                                                                                                                                                                                                                                                                                                                                                                                                                                                                                                                                                                                                                                                                                                                                                                                                                                                                                                                                                                                                                                                                                                                                                                                                                                                                                                                                                                                                                                                                                                                                                                                                                                                                                                                                                                                                                                                                                                                                                                                                                                                                                                                                                                                                                                                                                                                                                                                                                          
\usepackage{amsmath}
\usepackage{slashed}
\usepackage{anyfontsize}
\usepackage{bm}
\usepackage{slashed}
\usepackage{multirow}
\usepackage{float}
\usepackage[font=small]{caption}
\usepackage{listings}
\newgeometry{lmargin=1.1in, rmargin=1.1in, tmargin=1in, bmargin=1in}
\title{Generalized Abelian Gauge Field Theory under Rotor Model}
\author{B.T.T.Wong\footnote{CERN, u3500478@connect.hku.hk}}
\date{}

\begin{document}

\maketitle
\begin{abstract}
Gauge field theory with rank-one field $T_{\mu}$ is a quantum field theory that describes the interaction of elementary spin-1 particles, of which being massless to preserve gauge symmetry. In this paper, we give a generalized, extended study of abelian gauge field theory under successive rotor model in general $D$-dimensional flat spacetime for spin-1 particles in the context of higher order derivatives. We establish a theorem that $n$ rotor contributes to the $\Box^n T^{\mu}$ fields in the integration-by-parts formalism of the action. This corresponds to the transformation of gauge field $T^{\mu} \rightarrow \Box^n T^{\mu}$ and gauge field strength $G_{\mu\nu}\rightarrow \Box^n G_{\mu\nu} $ in the action. The $n=0$ case restores back to the standard abelian gauge field theory. The equation of motion and Noether's conserved current of the theory are also studied.
\end{abstract}

\section{Introduction}
Higher order derivative field theories have aroused interest in the study of quantum field theory because of their potential to eliminate infinities in scattering amplitudes \cite{ho1, ho2, ho3, ho4, ho5}. Yet, there are difficulties in the construction of a formalism due to the problem of renormalizability \cite{ho6, ho6a}. There are insights for higher derivative theories in quantum gravity and modified gravity \cite{h1,h2,h3, h4,h5} Numerous studies on higher order derivative field theories have been performed, including both scalar fields and gauge fields \cite{ho6a,ho6b,ho6c,ho8}.  However, these theories suffer from linear instability in dynamics, in which the energy function in the corresponding Hamiltonian would be unbounded below \cite{ho6d, ho6e, ho6f, ho6g, ho6h}. Nonetheless, higher order derivative field theories are essential to generalize the field theory of first or second order derivatives.  One of the examples is the extended Maxwell-Chern-Simons (MCS) Model in $D=2 +1$ three spacetime  dimension  \cite{ho7, ho8, ho9, ho10},
\begin{equation} \label{eq:chernSS}
    S= \int d^3 x \bigg( -\frac{1}{4}G_{\mu\nu}G^{\mu\nu} + \frac{g}{2}\epsilon^{\alpha\beta\gamma} (\Box T_{\alpha})(\partial_{\beta}T_{\gamma})   \bigg) \,,
\end{equation}
where $T_{\mu}$ is the $spin$-1 gauge field, $G_{\mu\nu}=\partial_{\mu}T_{\nu}-\partial_{\nu}T_{\mu}$ is the gauge field strength, $g$ is the coupling,  and the second term contains a higher order derivative term of $\Box T_{\alpha}$, contrasting to the normal Chern-Simons term of $\epsilon^{\alpha\beta\gamma}  T_{\alpha}\partial_{\beta}T_{\gamma} $. The extended MCS model is an example of a third-order derivative gauge field theory. 

Higher order derivative Maxwell abelian gauge field theory and its stability has been studied in J. Dai's paper using $n^{\mathrm{th}}$ order polynomial of the Maxwell operator \cite{ho6b}. In this article, we would like to construct a novel generalized higher order derivative gauge field theory, up to $2n^{\mathrm{th}}$ order, for the free abelian gauge field using the rotor model mechanism. 

For the first term in equation (1), it is the massless, free abelian gauge field theory of rank-one field $T_{\mu}$ that describes the dynamics of massless spin-1 particle, for example free photon in quantum electrodynamics (QED) process  \cite{GaugeField1, GaugeField2}. In this paper, we would intensively study this part of the action in $4$ dimesnion or higher with the context of higher order derivatives. The classical action for free abelian gauge field in 4D spacetime dimension is given by
\begin{equation} \label{eq:AbelianGaugeAction}
S = -\frac{1}{4} \int d^4 x G_{\mu\nu} G^{\mu\nu} \,.
\end{equation} 
By explicitly expanding equation (\ref{eq:AbelianGaugeAction}) together with integration by parts, and using the fact that the d'Alembertian $\Box = \partial_{\mu} \partial^{\mu}$ commutes with the Minkowski metric, $[ \Box , \eta_{\mu\nu} ] = 0 $, the action in (\ref{eq:AbelianGaugeAction}) is rewritten in the form as \cite{Peskin, Nair, Weinberg}
\begin{equation}
S = \frac{1}{2} \int d^4 x T^{\mu}(x) ( \Box \eta_{\mu\nu} - \partial_{\mu} \partial_{\nu}) T^{\nu} (x) = \frac{1}{2} \int d^4 x \, d^4 x^\prime \, \delta^4 (x-x^\prime) T^{\mu}(x) (\Box^\prime \eta^{\prime}_{\mu\nu} - \partial^{\prime}_{\mu} \partial^{\prime}_{\nu} ) T^{\nu}(x^\prime) \,.
\end{equation}
In field theory, the matrix element for the action is defined as \cite{Weinberg}
\begin{equation} \label{eq:MatrixElement}
M_{\mu\nu} (x, x^\prime) = -i \delta^4 (x-x^\prime) (\Box^\prime \eta^\prime_{ \,\mu \nu} - \partial^\prime_{\mu } \partial^\prime_{\nu} )   \,.
\end{equation}
In momentum space, the matrix element can be evaluated by inverse fourier transform,
\begin{equation}
M^{\mu\nu} (k , k^\prime) =  \int d^4 x d^4 x^\prime M^{\mu\nu}(x,x^\prime) e^{ikx}e^{ik^\prime x^\prime}\,.
\end{equation}
This will give us
\begin{equation}
M^{\mu\nu} (k , k^\prime) = -i (2\pi)^4 \delta^4 (k + k^\prime) (k^{\prime 2} \eta^{\prime} - k^{\prime \mu}k^{\prime \nu})\,.
\end{equation}
For convenience, we can always relabel $k$ as $k^\prime$, thus getting
\begin{equation}
M^{\mu\nu} (k , k^\prime) = -i (2\pi)^4 \delta^4 (k + k^\prime) (k^{ 2} \eta^{\mu\nu}- k^{ \mu}k^{\nu})\,.
\end{equation}

We would like to emphasize the tensorial part, which is a projector tensor. In the operator form and in the position space, we have the projection tensor as $\hat{R}_{\mu\nu} = \frac{1}{2}(\Box \eta_{\mu\nu} - \partial_{\mu}\partial_{\nu})$. In the momentum space we have $R_{\mu\nu} = \frac{1}{2}(k^2 \eta_{\mu\nu} - k_{\mu}k_{\nu})$. Therefore the matrix element of abelian gauge field theory in (\ref{eq:MatrixElement}) is proportional to the projection tensor. The action can hence be written as
\begin{equation}
S= \int d^4 x T^{\mu}\hat{R}_{\mu\nu} T^{\mu} =  \int d^4 k \tilde{T}^{\mu}(-k)R_{\mu\nu} \tilde{T}^{\nu}(k) \,,
\end{equation}
with the projection tensor coupling to two gauge fields. This allows us to define the Feynman vertex for free spin-1 boson as $i\int d^4 x \hat{R}_{\mu\nu}$ in the position space and $i\delta(k+k^\prime)R_{\mu\nu}$ in the momentum space.

The projection tensor can be thought as second-order rotation. It is useful to recall in the 3 dimensional case that
\begin{equation}
C_i = (\nabla \wedge (\nabla \wedge \vec{T}))_i = \epsilon_{ijk} \partial_j ( \epsilon_{klm}\partial_l T_m   ) = ( \delta_{il}\delta_{jm} - \delta_{im}\delta_{lj}  ) \partial_{j}\partial_{l} T_m = - (\nabla^2 \delta_{ij} - \partial_i \partial_j ) T_j \,.\\
\end{equation}
Therefore in 4 dimensional spacetime, similarly we have
\begin{equation}
C_{\mu}= ( \partial \wedge (\partial \wedge T))_{\mu} = - ( \Box \eta_{\mu\nu}- \partial_{\mu}\partial_{\nu}  ) T^{\nu} = -2\hat{R}_{\mu\nu} T^{\nu} \,.
\end{equation}
Next we define the rotor model of fields. The projector tensor acting on the gauge field can be regarded as second order rotation operation. For the first order, we have
\begin{equation}
L_{\mu} = \hat{R}_{\mu\nu}T^{\nu}\,.
\end{equation}
For the second order, we have
\begin{equation}
J^{\rho} = \hat{R}^{\rho\mu} \hat{R}_{\mu\nu}T^{\nu}\,,
\end{equation}
and so on. The number of times that the projection tensor acting on the gauge field is defined as the order of rotation in the rotor model. This contributes to a gauge field theory with higher order derivatives. The successive rotation acts as a source of generating higher order derivatives in the action.

\section{The generalized gauge field theorem under rotor field model}
\label{sec:1}
In this article, we aim at proving the following theorem, 
\begin{equation} \label{eq:general}
S = -\frac{1}{4} \int d^D x G_{n\,\mu\nu} G^{\mu\nu}_n  = \frac{1}{4^n}\int d^D x \big( \Box^n T^{\mu} \big) \hat{R}_{\mu\nu}  \big( \Box^n T^{\nu} \big) =- \frac{1}{4^{n+1}} \int d^{D}x \,\Box^n G_{\mu \nu} \Box^n G^{\mu \nu}\,,
\end{equation} 
where $G_{n\,\mu\nu}$ is the field strength of the $n^\mathrm{th}$ order rotor gauge field strength, $D$ is the general $D$ dimensional spacetime. The vertex tensor remains as the projector tensor in any order of second-ordered rotation. When $n=0$, this returns back to the standard gauge field theory,
\begin{equation} \label{eq:standard}
S = -\frac{1}{4} \int d^D x G_{\mu\nu} G^{\mu\nu} = \int d^D x T^{\mu}(x) \hat{R}_{\mu\nu} T^{\nu} (x) \,.
\end{equation}
Hence under the rotor model of successive rotation of gauge fields, it changes $T^{\mu}$ to $\Box ^n T^{\mu}$ oscillation field, with the vertex tensor being unchanged. 

The proof of the theorem would be conducted in the following way. First we prove the $n=1$ case, followed by $n=2$ case then the general $n^\mathrm{th}$ order case for the equivalence of the first two equations in (\ref{eq:general}).  Then we will prove the equivalence of the second and third equation in (\ref{eq:general}).

\subsection{The n=1 case}
\label{sec:2}
We begin with the first-ordered rotor field $L_{\mu} = \hat{R}_{\mu\nu}T^{\nu}$. The new field strength is
\begin{equation}
H_{\mu\nu} = \partial_{\mu} L_{\nu}- \partial_{\nu}L_{\mu} \,.
\end{equation}
The new action is
\begin{equation}
S= -\frac{1}{4} \int d^D x H_{\mu\nu} H^{\mu\nu} = -\frac{1}{2} \int d^D x \Big(\partial_{\mu} L_{\nu} \partial^{\mu} L^{\nu}   - \partial_{\mu}L_{\nu}  \partial^{\nu} L^{\mu} \Big)\,.  
\end{equation}
Then
\begin{equation}
\begin{aligned}
&\quad \partial_{\mu} L_{\nu} \partial^{\mu} L^{\nu} \\
&= \frac{1}{4} (\eta_{\nu\rho} \Box \partial_\mu T^\rho - \partial_{\mu} \partial_{\nu} \partial_{\rho} T^\rho ) ( 
\eta^{\nu\rho} \Box \partial^\mu T_\rho - \partial^{\mu}\partial^{\nu} \partial^{\rho} T_\rho ) \\
&= \frac{1}{4}\bigg( (\Box \partial_{\mu} T^{\sigma}) (\Box \partial^{\mu}T_{\sigma}) -2 (\partial^\mu \partial_{\rho} \partial^{\sigma}T_{\sigma}) \Box \partial_{\mu} T^{\rho}  +   (\partial_{\mu}  \partial_{\nu} \partial_{\rho} T^\rho) ( \partial^{\mu} \partial^{\nu} \partial^{\sigma} T_{\sigma}  ) \bigg)\,.\\
\end{aligned}
\end{equation}
Now we will carry out integration by parts for each term, and note that the boundary term vanishes.
The first term gives
\begin{equation}
\int d^D x \, (\Box \partial_{\mu} T^{\sigma}) (\Box \partial^{\mu}T_{\sigma})
 = - \int d^D x \, \Box T^{\sigma} ( \partial_{\mu}\Box \partial^{\mu} T_{\sigma})= - \int d^D x \,\Box T^{\sigma}\, \Box^2 T_{\sigma}\,.
\end{equation}
The second term gives
\begin{equation}
 -2 \int d^D x \, (\partial^\mu \partial_{\rho} \partial^{\sigma}T_{\sigma}) \Box \partial_{\mu} T^{\rho} = 2 \int d^D x  \,\Box T^{\rho} ( \partial_{\mu}\partial^{\mu} \partial_\rho \partial^{\sigma}T_{\sigma})=  2 \int d^D x \,  \Box  T^{\rho} \,\Box \partial_\rho \partial^{\sigma}T_{\sigma} \,.
\end{equation}
For the third term we need to do integration by parts for three times,
\begin{equation}
\begin{aligned}
&\quad \int d^D x \,(\partial_{\mu}  \partial_{\nu} \partial_{\rho} T^\rho) ( \partial^{\mu} \partial^{\nu} \partial^{\sigma} T_{\sigma}  ) \\
& = -\int d^D x \,\partial^{\nu} \partial^{\sigma} T_{\sigma} \,\,\Box \partial_{\nu}\partial_{\rho}T^{\rho} \\
&= + \int d^D x \,\Box \partial_{\rho}T^{\rho}( \Box \partial_{\rho} \partial^{\sigma}T_{\sigma})\\
& = -\int d^D x \,\Box T^{\rho} ( \Box  \partial_{\rho} \partial^{\sigma} T_{\sigma} )\,.\\
\end{aligned}
\end{equation}
Therefore the first term of the action is 
\begin{equation} \label{eq:FirstTermAction}
-\frac{1}{2} \int d^D x  \,\partial_{\mu} L_{\nu} \partial^{\mu} L^{\nu}  = \frac{1}{8}\int d^D x \,\Box T^{\sigma} \Big(\Box^2 T_{\sigma} - \Box \partial_{\sigma} \partial^{\rho}T_{\rho} \Big) = \frac{1}{8} \int d^D x \,\Box T^{\sigma} \Big( \eta_{\sigma\rho} \Box - \partial_{\sigma}\partial_{\rho} \Big) \,\Box T^{\rho}\,.
\end{equation}
Then we evaluate the second term of the action.
\begin{equation}
\begin{aligned}
&\quad \partial_{\mu} L_{\nu} \partial^{\nu} L^{\mu} \\
&= \frac{1}{4} (\eta_{\nu\rho} \Box \partial_\mu T^\rho - \partial_{\mu} \partial_{\nu} \partial_{\rho} T^\rho ) ( 
\eta^{\mu\alpha} \Box \partial^\nu T_\alpha - \partial^{\nu}\partial^{\mu} \partial^{\alpha} T_\alpha ) \\
&= \frac{1}{4}\bigg( \eta_{\nu\rho}\eta^{\mu\alpha} \Box\partial_{\mu} T^{\rho}\,\Box \partial_{\nu}T_{\alpha} -2\Box \partial_{\mu} T^{\rho} (\partial_{\rho}\partial^{\mu}\partial^{\alpha}T_{\alpha} ) +  ( \partial_{\mu}\partial_{\nu}\partial_{\rho}T^\rho )( \partial^{\mu} \partial^{\nu} \partial_{\alpha}T^{\alpha}) \bigg) \\
&= \frac{1}{4} \bigg((\partial^\alpha \Box T^{\rho}) (\partial_{\rho}\Box T_{\alpha})-2\Box \partial_{\mu} T^{\rho} (\partial_{\rho}\partial^{\mu}\partial^{\alpha}T_{\alpha} ) +  ( \partial_{\mu}\partial_{\nu}\partial_{\rho}T^\rho )( \partial^{\mu} \partial^{\nu} \partial_{\alpha}T^{\alpha}) \bigg)\,. \\
\end{aligned}
\end{equation}
Again we carry out integration by parts for each term.
For the first term
\begin{equation}
\int d^D x (\partial^\alpha \Box T^{\rho}) (\partial_{\rho}\Box T_{\alpha})= - \int d^D x \,\Box T^{\rho} \Big( \Box \partial_{\rho}\partial^{\alpha}T_{\alpha}  \Big)\,.
\end{equation}
The second term is
\begin{equation}
-2\int d^D x \Box \partial_{\mu} T^{\rho} (\partial_{\rho}\partial^{\mu}\partial^{\alpha}T_{\alpha} ) = 2\int d^D x \Box T^{\rho} \Big(\partial_{\mu} \partial_{\rho}\partial^{\mu}\partial^{\alpha} T_{\alpha}\Big) =2\int d^D x \Box T^{\rho} \Big( \Box \partial_{\rho}\partial^{\alpha}T_{\alpha} \Big) \,.
\end{equation}
The third term is just same as the one in the last term in the first action
\begin{equation}
-\int d^D x \,\Box T^{\rho} \Big( \Box \partial_{\rho} \partial^{\alpha}T_{\alpha} \Big)\,.
\end{equation} 
Therefore we find the second term in the action cancels,
\begin{equation}
-\frac{1}{2}\int d^D x \,\Box T^{\rho} \Big( -\Box \partial_{\rho} \partial^{\alpha}T_{\alpha} + 2 \Box \partial_{\rho} \partial^{\alpha}T_{\alpha} - \Box \partial_{\rho} \partial^{\alpha}T_{\alpha}  \Big) =0 \,.\\
\end{equation}
Hence the action for the first-order rotated field is
\begin{equation}
H= -\frac{1}{4} \int d^D x H_{\mu\nu} H^{\mu\nu} = \frac{1}{8} \int d^D x \Box T^{\sigma} \Big( \eta_{\sigma\rho} \Box - \partial_{\sigma}\partial_{\rho} \Big) \Box T^{\rho} = \frac{1}{4} \int d^D x  \Box T^{\mu} \hat{R}_{\mu\nu} \Box T^{\nu}\,.
\end{equation}
This completes the proof of the $n=1$ case.

\subsection{The n=2 case}
\label{sec:3}
Now we would like to continue to construct a new rotation gauge field from $L_{\mu}$, meaning that we apply the projection tensor twice on our original vector field $T^\nu$. We get
\begin{equation}
J^{\rho} = \hat{R}^{\rho\mu}L_{\mu} = \hat{R}^{\rho\mu} \hat{R}_{\mu\nu}T^{\nu}\,,
\end{equation}
and we would like to construct another gauge field strength  $K_{\mu\nu} = \partial_{\mu}J_{\nu} - \partial_{\nu}J_{\mu}$. First we compute the product of the two projection tensors
\begin{equation}
\begin{aligned}
\hat{R}^{\rho\mu}\hat{R}_{\mu\nu} & = \frac{1}{4} ( \Box \eta^{\rho\mu} - \partial^{\rho} \partial^{\mu} ) (\Box \eta_{\mu\nu} - \partial_{\mu} \partial_{\nu})\\
& = \frac{1}{4}  \Big( \Box^2 \delta^{\rho}_{\nu} -2\Box \eta^{\rho\mu} \partial_{\mu}\partial_{\nu} +  \partial^{\rho} \Box \partial_{\nu}   \Big)\\
&= \frac{1}{4} \Big(\Box^2 \delta^{\rho}_{\nu} - \Box \partial^{\rho}\partial_{\nu} \Big) =\frac{1}{4}\Box \Big( \Box   \delta^{\rho}_{\nu} -  \partial^{\rho}\partial_{\nu}\Big)\,.\\
\end{aligned}
\end{equation}
Therefore, we have the gauge field as follow,
\begin{equation}
J^{\mu} = \frac{1}{4} \Big( \Box^2 \delta^{\mu}_{\alpha} - \Box \partial^{\mu}\partial_{\alpha} \Big)T^{\alpha}\,.
\end{equation}
And we construct the action as
\begin{equation}
S= -\frac{1}{4} \int d^D x  K_{\mu\nu} K^{\mu\nu} = -\frac{1}{2} \int d^D x \Big(\partial_{\mu}J_{\nu} \partial^{\mu} J^{\nu} - \partial_{\mu} J_{\nu} \partial^{\nu}J^{\mu}  \Big)\,.
\end{equation}
Consider the first term of the Lagranigan in the action,
\begin{equation}
\begin{aligned}
&\quad \partial_{\mu}J_{\nu} \partial^{\mu} J^{\nu} \\
&= \frac{1}{16}\Big( \Box^2 \partial_{\mu} \delta^{\alpha}_{\nu} - \Box \partial_{\mu} \partial_{\nu} \partial^{\alpha} \Big) T_{\alpha} \Big( \Box^2 \partial^{\mu} \delta^{\nu}_{\beta} - \Box\partial^{\mu} \partial^{\nu} \partial_{\beta} \Big)T^{\beta}\\
&= \frac{1}{16} \Big( (\Box^2 \partial_{\mu} T_{\nu})(\Box^2 \partial^{\mu} T^{\nu})  -2(\Box^2 \partial_{\mu} T_{\nu} )(\Box \partial^{\mu} \partial^{\nu} \partial_{\beta} T^{\beta} ) +  ( \Box  \partial_{\mu} \partial_{\nu} \partial^{\alpha}T_{\alpha}) (\Box \partial^{\mu} \partial^{\nu} \partial_{\beta} T^{\beta})  \Big) \,.\\
\end{aligned}
\end{equation}
Next we evaluate each term using intergration by parts. The first term gives
\begin{equation}
\int d^D x (\Box^2 \partial_{\mu} T_{\nu})(\Box^2 \partial^{\mu} T^{\nu})  = - \int d^D x (\Box^2  T_{\nu}) (\partial_{\mu} \Box^2  \partial^{\mu} T^{\nu}) = -\int d^D x \Box^2 T_{\nu} (\Box^3 T^{\nu})\,.
\end{equation}
The second term gives
\begin{equation}
-2 \int d^D x  ( \Box^2 \partial_{\mu}T_{\nu} ) (\Box \partial^{\mu} \partial^{\nu} \partial_{\beta} T^{\beta} ) = 2 \int d^D x \Box^2 T_{\nu} (\Box^2 \partial^{\nu} \partial_{\beta} T^{\beta}  )\,.
\end{equation}
For the third term we perform integration by parts for three times on $\partial_{\mu} , \partial_{\nu} , \partial^{\alpha}$ and this gives
\begin{equation} \label{eq:JThirdTerm}
\int d^D x ( \Box  \partial_{\mu} \partial_{\nu} \partial^{\alpha}T_{\alpha}) (\Box \partial^{\mu} \partial^{\nu} \partial_{\beta} T^{\beta})   = -\int d^D x \,(\Box T_{\alpha}) (\Box^3 \partial^{\alpha}\partial_{\beta} T^\beta )\,.
\end{equation}
Therefore the full first term of the action is 
\begin{equation} \label{eq:JActionFirstTerm}
-\frac{1}{2} \int d^D x \,\partial_{\mu}J_{\nu} \partial^{\mu} J^{\nu} = \frac{1}{32} \int d^D x \Big( (\Box^2 T_{\nu}) (\Box^3 T^{\nu}) -2 (\Box^2 T_{\nu}) (\Box^2 \partial^{\nu} \partial_{\beta} T^{\beta} )+  (\Box T_{\alpha}) (\Box^3 \partial^{\alpha}\partial_{\beta} T^\beta ) \Big) \,.
\end{equation}
It noted that unlike the previous $L_{\mu}$ case, here for $J_{\mu}$ the last two terms of the first-term action do not cancel. Next we evaluate the second Lagrangian term of the action
\begin{equation}
\begin{aligned}
&\quad \partial_{\mu}J_{\nu} \partial^{\nu} J^{\mu} \\
&= \frac{1}{16} \Big( (\Box^2 \partial_{\mu} T_{\nu})(\Box^2 \partial^{\nu} T^{\mu})  -2(\Box \partial_{\mu} \partial_{\nu} \partial^{\alpha} T_{\alpha}) (\Box^2 \partial^{\nu} T^{\mu} ) +  ( \Box  \partial_{\mu} \partial_{\nu} \partial^{\alpha}T_{\alpha}) (\Box \partial^{\mu} \partial^{\nu} \partial_{\beta} T^{\beta})  \Big) \,.\\
\end{aligned}
\end{equation}
Then we  evaluate each term for the second-term action using integration by parts. The first term gives,
\begin{equation}
\int d^D x (\Box^2 \partial_{\mu} T_{\nu})(\Box^2 \partial^{\nu} T^{\mu}) = -\int d^D x \Box^2 T_{\nu} (\Box^2 \partial^{\nu} \partial_{\mu} T^{\mu} )\,.
\end{equation}
The second term gives,
\begin{equation}
-2 \int d^D x (\Box \partial_{\mu} \partial_{\nu} \partial^{\alpha} T_{\alpha}) (\Box^2 \partial_{\nu} T_{\mu} ) = 2\int d^D x (\Box^2 T^{\mu})  (\Box^2 \partial_{\mu} \partial^{\alpha}T_{\alpha})\,.
\end{equation}
The third term will be same as the third term in (\ref{eq:JThirdTerm}). Hence the full action of the second term is 
\begin{equation} \label{eq:JActionSecondTerm}
-\frac{1}{2} \int d^D x \,\partial_{\mu}J_{\nu} \partial^{\nu} J^{\mu}
=\frac{1}{32} \int d^D x \Big( (\Box^2 T_{\nu}) (\Box^2 \partial^{\nu} \partial_{\mu} T^{\mu} ) - 2 (\Box^2 T^{\nu} )(\Box^2  \partial_{\nu} \partial^{\alpha} T_{\alpha}) + (\Box T_{\alpha} )(\Box^3 \partial^{\alpha} \partial_{\beta} T^{\beta} ) \Big)\,.
\end{equation}
Finally the full action is given by (\ref{eq:JActionFirstTerm}) minus (\ref{eq:JActionSecondTerm}), where we can see the last two terms canceling each other. Thus we obtain
\begin{equation}
\begin{aligned}
S &= -\frac{1}{4} \int d^D x K_{\mu\nu} K^{\mu\nu} \\
&= \frac{1}{32}\int d^D x \Big( (\Box^2 T_{\nu})(\Box^3 T^{\nu}) - (\Box^2 T_{\nu}) (\Box^2 \partial^{\nu} \partial_{\mu}T^{\mu} )\Big)\\
& =  \frac{1}{32}\int d^D x   (\Box^2 T_{\nu})\Big(\Box^3 T^{\nu} -\Box^2 \partial^{\nu}\partial_{\mu} T^{\mu} \Big)\,,\\
\end{aligned}
\end{equation}
as we have $\Box^3 T_{\nu} = \Box (\Box^2 T_{\nu}) = \eta_{\mu\nu}\Box (\Box^2 T^{\mu})$ and the Minkowski metric tensor commutes with the box operator. Therefore we obtain 
\begin{equation}
S = \frac{1}{32} \int d^D x(\Box^2 T^{\nu} ) \Big( \Box \eta_{\mu\nu} -\partial_{\mu}\partial_{\nu}\Big) ( \Box^2 T^{\mu}) \,,
\end{equation}
then by swapping the dummy indices $\mu$ and $\nu$ we obtain the final result as
\begin{equation}
\begin{aligned}
S = -\frac{1}{4} \int d^D x K_{\mu\nu} K^{\mu\nu} &=\frac{1}{32} \int d^D x (\Box^2 T^{\mu} ) \Big( \Box \eta_{\mu\nu} -\partial_{\mu}\partial_{\nu}\Big)( \Box^2 T^{\nu} ) \\
& =\frac{1}{16}\int d^D x  (\Box^2 T^{\mu} ) \hat{R}_{\mu\nu}( \Box^2 T^{\nu} ) \,.
\end{aligned}
\end{equation}
Thus this completes the proof for $n=2$ case.

\subsection{The general n case}
\label{sec:4}
Using the results from the $n=1$ and the $n=2$ case, we can promote to prove the general $n$ case.
We would like to show that this is generally true for all cases when we continue to act on the original $T_{\mu}$ compositely by the second-ordered rotation operation. For convenience we will introduce a systematic way for index labelling. Firstly we recall that
\begin{equation}
\hat{R}_{\mu_1 \mu_0} = \frac{1}{2} \Big( \Box \eta_{\mu_1  \mu_0} - \partial_{\mu_1} \partial_{\mu_0}\Big)\,,
\end{equation}
is a rank (0,2) tensor (which has no upper indices and has two lower indices). For even number of rotations, we will get a $(1,1)$ rank tensor (which has one upper index and one lower index), we define $C^{\mu_2}_{\,\,\,\mu_0}$ as
\begin{equation}
C^{\mu_2}_{\,\,\,\mu_0} = \hat{R}^{\mu_2 \mu_1} \hat{R}_{\mu_1 \mu_0} = \frac{1}{4} \Big( \Box^2 \delta^{\mu_2}_{\,\,\,\mu_0} - \Box \partial^{\mu_2} \partial_{\mu_0} \Big) = \frac{1}{4} \Box \Big( \Box \delta^{\mu_2}_{\,\,\,\mu_0} - \partial^{\mu_2} \partial_{\mu_0} \Big)\,.
\end{equation}
We can obtain a rank (0,2) tensor by pulling down the upper index by the flat Minkowski metric tensor, then we have
\begin{equation}
C_{\rho_2 \mu_0} = \eta_{\rho_2 \mu_2} \hat{R}^{\mu_2 \mu_1} \hat{R}_{\mu_1 \mu_0}  = \frac{1}{4} \Box \Big( \Box \eta_{\rho_2 \mu_0} - \partial_{\rho_2} \partial_{\mu_0}  \Big) = \frac{1}{2} \Box \hat{R}_{\rho_2 \mu_0} =\frac{1}{2} (\Box \delta^{\mu_1}_{\,\,\,\rho_2} )\hat{R}_{\mu_1 \mu_0} \,.
\end{equation}
Hence second-ordered rotation can be considered as the oscillation of the first rotation. This shows how a rotation is equivalent to an oscillation, given that the appearance of the d'Alembert operator. We define the propagator \footnote{Here the term propagator is used and should not be confused with the normal Feynman propagator used in quantum field theory. } as,
\begin{equation}
P^{\mu_1}_{\,\,\,\rho_2} = \Box \delta^{\mu_1}_{\,\,\,\rho_2} \,.
\end{equation}
Now we proceed to three rotations,
\begin{equation}
\begin{aligned}
&\quad \hat{R}_{\mu_3 \mu_2} \hat{R}^{\mu_2 \mu_1} \hat{R}_{\mu_1 \mu_0} \\
&= \frac{1}{2} \Big(\Box  \eta_{\mu_3 \mu_2} -\partial_{\mu_3} \partial_{\mu_2}\Big) \cdot \frac{1}{4}\Big(\Box^2 \delta^{\mu_2}_{\mu_0} - \Box \partial^{\mu_2} \partial_{\mu_0}  \Big)\\
&= \frac{1}{8} \Big( \Box^3 \eta_{\mu_3 \mu_0} - 2 \Box^2 \partial_{\mu_3} \partial_{\mu_0} + \partial_{\mu_3} \Box^2 \partial_{\mu_0} \Big) \\
&= \frac{1}{8} \Box^2 \Big( \Box \eta_{\mu_3 \nu_0} -  \partial_{\mu_3} \partial_{\mu_0}  \Big)\\
&= \frac{1}{4} \Box^2 \hat{R}_{\mu_3 \mu_0} = \frac{1}{4} \Big( \Box^2 \delta^{\mu_1}_{\,\,\,\mu_3} \Big) \hat{R}_{\mu_1 \mu_0} \,.\\
\end{aligned}
\end{equation}
Thus we define the propagator as
\begin{equation}
P^{\mu_1}_{\,\,\,\mu_3} =\Box^2 \delta^{\mu_1}_{\,\,\,\mu_3} = (\Box \delta^{\mu_1}_{\,\,\,\mu_2})(\Box \delta^{\mu_2}_{\,\,\,\mu_3}) =  P^{\mu_1}_{\,\,\,\mu_2} P^{\mu_2}_{\,\,\,\mu_3}\,,
\end{equation}
which acts on the projection tensor. Next we proceed to four rotations, 
\begin{equation}
\begin{aligned}
&\quad \hat{R}^{\mu_4 \mu_3}\hat{R}_{\mu_3 \mu_2} \hat{R}^{\mu_2 \mu_1} \hat{R}  _{\mu_1 \mu_0} = C^{\mu_4}_{\,\,\,\mu_2} C^{\mu_2}_{\,\,\,\mu_0}\\
& =\frac{1}{4} \Big( \Box^2 \delta^{\mu_4}_{\,\,\,\mu_2}- \Box \partial^{\mu_4} \partial_{\mu_2} \Big) \cdot \frac{1}{4} \Big( \Box^2 \delta^{\mu_2}_{\,\,\,\mu_0}- \Box \partial^{\mu_2} \partial_{\mu_0} \Big)\\
& =\frac{1}{16} \Big( \Box^4 \delta^{\mu_4}_{\mu_2} \delta^{\mu_2}_{\mu_0} - 2 \Box^3 \partial^{\mu_4} \partial_{\mu_0} + \Box^3   \partial^{\mu_4} \partial_{\mu_0}\Big)\\
&= \frac{1}{16} \Box^3 \Big(\Box \delta^{\mu_4}_{\,\,\,\mu_0} -  \partial^{\mu_4} \partial_{\mu_0} \Big) \,. \\
\end{aligned}
\end{equation}
And again we can pull down the index with the metric tensor,
\begin{equation}
C_{\rho_{4} \mu_0} = \eta_{\rho_4 \mu_4}  C^{\mu_4}_{\,\,\,\mu_2} C^{\mu_2}_{\,\,\,\mu_0} = \frac{1}{8} \Box^3 \hat{R}_{\rho_4 \mu_0} = \frac{1}{8} \Big( \Box^3 \delta^{\,\,\,\,\mu_1}_{ \rho_4} \Big)\hat{R}_{\mu_1 \mu_0}\,.
\end{equation}
The propagator is defined as
\begin{equation}
P_{\rho_4}^{\,\,\,\,\mu_1} = (\Box \delta_{\rho_4}^{\,\,\,\,\rho_3} )(\Box \delta_{\rho_3}^{\,\,\,\,\rho_2}   ) (\Box \delta_{\rho_2}^{\,\,\,\,\mu_1} ) \,.
\end{equation}
It can be observed that due to the nature of difference between upper and lower indices, the order of rotations is separated into odd and even parts. This will not occur if all the indices are in lower (or upper) positions. But raising or lowering indices arise from the fact that we are working in a Minkowski manifold such that it allows negative metric tensor components. The above has shown the result for rotation once and twice, we have also shown that these rotations can be represented by composite d'alembert operations. Now we would like to generalize the result to $n$ cases.

Let $n$ be the number of rotations defined by $n$-times action of the projection tensor. For $n=2k-1$ is odd,
\begin{equation}
\begin{aligned}
 \hat{R}_{\mu_n \mu_{n-1}}\hat{R}^{\mu_{n-1} \mu_{n-2}} \cdots \hat{R}_{\mu_3 \mu_2} \hat{R}^{\mu_2 \mu_1} \hat{R}_{\mu_1 \mu_0} &= \frac{1}{2^{n-1}} \Big( \Box \eta_{\mu_n \mu_{n-1}} \Box \eta^{\mu_{n-1} \mu_{n-2}} \cdots \Box \eta^{\mu_2 \mu_1}  \Big) \hat{R}_{\mu_1 \mu_0} \\
 &= \frac{1}{2^{n-1}} \Big( \Box \delta_{\mu_n}^{\,\,\,\mu_{n-1}}  \Box \delta_{\mu_{n-1}}^{\,\,\,\mu_{n-2}}\cdots \Box\delta_{\mu_3}^{\,\,\,\mu_2} \Box \delta_{\mu_1}^{\,\,\,\mu_2} \Big) \hat{R}_{\mu_1 \mu_0} \\
&= \frac{1}{2^{n-1}}\Big(\Box^{n-1} \delta_{\mu_n}^{\,\,\,\mu_1} \Big)\hat{R}_{\mu_1 \mu_0}\,.\\
\end{aligned}
\end{equation}
For $n=2k$ is even,
\begin{equation}
\begin{aligned}
\hat{R}^{\mu_n \mu_{n-1}}\hat{R}_{\mu_{n-1} \mu_{n-2}} \cdots \hat{R}_{\mu_3 \mu_2} \hat{R}^{\mu_2 \mu_1} \hat{R}_{\mu_1 \mu_0} &= \frac{1}{2^n} \Big( \Box \eta^{\mu_n \mu_{n-1}} \Box \eta_{\mu_{n-1} \mu_{n-2}} \cdots \Box \eta^{\mu_2 \mu_1}  \Big) \hat{R}_{\mu_1 \mu_0} \,,
\end{aligned}
\end{equation}
and with the lowered indices as 
\begin{equation}
\begin{aligned}
C_{\rho_n \mu_0} &= \eta_{\rho_n \mu_n}\hat{R}^{\mu_n \mu_{n-1}}\hat{R}_{\mu_{n-1} \mu_{n-2}} \cdots \hat{R}_{\mu_3 \mu_2} \hat{R}^{\mu_2 \mu_1} \hat{R}_{\mu_1 \mu_0}\\ 
&= \frac{1}{2^{n-1}} \Big( \Box \delta_{\rho_n}^{\,\,\,\mu_{n-1}}  \Box \delta_{\mu_{n-1}}^{\,\,\,\mu_{n-2}} \cdots \Box\delta_{\mu_3}^{\,\,\,\mu_2} \Box \delta_{\mu_2}^{\,\,\,\mu_1} \Big) \hat{R}_{\mu_1 \mu_0}\\
&= \frac{1}{2^{n-1}}\Big(\Box^{n-1} \delta_{\rho_n}^{\,\,\,\mu_1} \Big)\hat{R}_{\mu_1 \mu_0}\,.\\
\end{aligned}
\end{equation}
Thus we can synchronize both the odd and even cases as rank (0,2) tensors.
We define formally the propagators as \begin{equation}
P_{\mu_j}^{\,\,\,\mu_{j-1}} = \Box \delta_{\mu_j}^{\,\,\,\mu_{j-1}}\,.
\end{equation}
Then successive rotations can always be considered as the propagations of the first rotation. For example in the odd case, we express in terms of propagators as
\begin{equation}
\hat{R}_{\mu_n \mu_{n-1}}\hat{R}^{\mu_{n-1} \mu_{n-2}} \cdots \hat{R}_{\mu_3 \mu_2} \hat{R}^{\mu_2 \mu_1} \hat{R}_{\mu_1 \mu_0} = \frac{1}{2^{n-1}}  P_{\mu_n}^{\,\,\,\mu_{n-1}} P_{\mu_{n-1}}^{\,\,\,\mu_{n-2}} \cdots P_{\mu_3}^{\,\,\,\mu_{2}} P_{\mu_2}^{\,\,\,\mu_{1}} \hat{R}_{\mu_1 \mu_0} \,,
\end{equation}
similarly for the even case. Next, we define the set of gauge fields which are generated by the successive second-ordered rotations of the projection tensor, $\{ T_{n}^{\mu} , T_{n-1}^{\mu} , \cdots , T_{1}^{\mu} , T_{0}^{\mu} \}$ (lower indices for odd and upper indices for even). The action of the gauge field formed by $n$-rotation operation is 
\begin{equation}
S= - \frac{1}{4} \int d^D x G_{n\,\mu\nu} G_n^{\mu\nu} = -\frac{1}{2} \int d^D x \Big( \partial_{\mu} T_{n\, \nu} \partial^{\mu} T_{n}^{\nu} -   \partial_{\mu} T_{n\, \nu} \partial^{\nu} T_{n}^{\mu}  \Big)\,,
\end{equation}
where the field strength is 
\begin{equation}
G_{n\, \mu\nu} = \partial_{\mu} T_{n\,\nu}- \partial_{\nu} T_{n \,\mu}\,.
\end{equation}
The $n$-th second-ordered rotated field is, for $n$ is odd;
\begin{equation}
T_{\mu_n} = (  \hat{R}_{\mu_n \mu_{n-1}}\hat{R}^{\mu_{n-1} \mu_{n-2}} \cdots \hat{R}_{\mu_3 \mu_2} \hat{R}^{\mu_2 \mu_1} \hat{R}_{\mu_1 \mu_0}  )T^{\mu_0} = \frac{1}{2^{n-1}}\Big(\Box^{n-1} \delta_{\mu_n}^{\,\,\,\mu_1} \Big)\hat{R}_{\mu_1 \mu_0}T^{\mu_0} \,, 
\end{equation}
and for $n$ is even;
\begin{equation}
T_{\rho_n} = (  \eta_{\rho_n \mu_n}\hat{R}^{\mu_n \mu_{n-1}}\hat{R}_{\mu_{n-1} \mu_{n-2}} \cdots \hat{R}_{\mu_3 \mu_2} \hat{R}^{\mu_2 \mu_1} \hat{R}_{\mu_1 \mu_0}  )T^{\mu_0} = \frac{1}{2^{n-1}}\Big(\Box^{n-1} \delta_{\rho_n}^{\,\,\,\mu_1} \Big)\hat{R}_{\mu_1 \mu_0}T^{\mu_0} \,, 
\end{equation}
When we apply these to the action, it does not matter whether it is the odd case or the even case as summing over dummy indices will automatically take account into both cases.

Consider the first Lagrangian term in the action, and relabel all indices with subscript $n$,
\begin{equation}
\begin{aligned}
\partial_{\alpha_n} T_{\mu_n} \partial^{\alpha_n} T^{\mu_n}& = \frac{1}{4^{n-1}} \Big[ \Box^{n-1} (\delta^{\,\,\,\mu_1}_{\mu_n} \partial_{\alpha_n} R_{\mu_1 \mu_0})T^{\mu_0} \Big] \Big[ \Box^{n-1} ( \delta_{\sigma_1}^{\,\,\, \mu_n} \partial^{\alpha_n} \hat{R}^{\sigma_1 \sigma_0}) T_{\sigma_0} \Big]\\
&= \frac{1}{4^{n-1}} \Big[ \Box^{n-1} (\partial_{\alpha_n} R_{\mu_n \mu_0} T^{\mu_0}) \Big] \Big[ \Box^{n-1} (\partial^{\alpha_n} R^{\mu_n \sigma_0} T_{\sigma_0}) \Big] \,.\\
\end{aligned}
\end{equation}
Now we substitute the explicit form of the first projection tensor,
\begin{equation}
\hat{R}_{\mu_1 \mu_0} = \frac{1}{2} \Big( \Box \eta_{\mu_1 \mu_0} - \partial_{\mu_1} \partial_{\mu_0} \Big) \,,
\end{equation}
then the first Lagrangian term is 
\begin{equation}
\begin{aligned}
& \quad \partial_{\alpha_n} T_{\mu_n} \partial^{\alpha_n} T^{\mu_n}\\
&= \frac{1}{ 4^n} \Big[ \Box^{n-1} \partial_{\alpha_n} \Big( \Box \eta_{\mu_n \mu_0} - \partial_{\mu_n} \partial_{\mu_0} \Big) T^{\mu_0} \Big] \Big[  \Box^{n-1} \partial^{\alpha_n} \Big( \Box \eta^{\mu_n \sigma_0} - \partial^{\mu_n} \partial^{\sigma_0} \Big) T_{\sigma_0} \Big] \\
&= \frac{1}{ 4^n} \Big[ \Big( \Box^n  \partial_{\alpha_n}T_{\mu_n} \Big)   \Big( \Box^n  \partial^{\alpha_n}T^{\mu_n} \Big) - \Big( \Box^n \partial_{\alpha_n} T_{\mu_n} \Big) \Big( \Box^{n-1} \partial^{\alpha_n} \partial^{\mu_n} \partial^{\sigma_0} T_{\sigma_0} \Big)\\
&\quad  - \Big( \Box^{n-1} \partial_{\alpha_n} \partial_{\mu_n} \partial_{\mu_0} T^{\mu_0} \Big) \Big( \Box^{n} \partial^{\alpha_n} T^{\mu_n}\Big) + \Big( \Box^{n-1} \partial_{\alpha_n} \partial_{\mu_n} \partial_{\mu_0} T^{\mu_0}\Big)   \Big(\Box^{n-1} \partial^{\alpha_n} \partial^{\mu_n} \partial^{\sigma_0} T_{\sigma_0} \Big) \Big] \,.\\
\end{aligned}
\end{equation}
Therefore the first action term is:
\begin{equation}
\begin{aligned}
&\quad -\frac{1}{2  } \int d^D x \big( \partial_{\mu} T_{n\, \nu} \partial^{\mu} T_{n}^{\nu} \big) \\
&= -\frac{1}{2^{2n +1}} \int d^D x  \big[ \big( \Box^n  \partial_{\alpha_n}T_{\mu_n} \big) \big( \Box^n \partial^{\alpha_n} T^{\mu_n}\big) -2 \big( \Box^n \partial_{\alpha_n} T_{\mu_n}   \big) \big( \Box^{n-1} \partial^{\alpha_n} \partial^{\mu_n} \partial^{\sigma_0} T_{\sigma_0} \big) \\
& \quad +\big(\Box^{n-1} \partial_{\alpha_n} \partial_{\mu_n} \partial_{\mu_0} T^{\mu_0} \big) \big( \Box^{n-1}  \partial^{\alpha_n} \partial^{\mu_n} \partial^{\sigma_0} T_{\sigma_0} \big)\big]  \,.\\
\end{aligned}
\end{equation}
Next we will perform integration by parts for each term. The first term gives
\begin{equation}
\int d^D x \big( \Box^n  \partial_{\alpha_n}T_{\mu_n} \big) \big( \Box^n  \partial^{\alpha_n}T^{\mu_n} \big) = - \int d^D x \big(\Box^n T_{\mu_n} \big) \big(\Box^{n+1} T^{\mu_n} \big)\,.
\end{equation}
The second term gives
\begin{equation}
-2 \int d^D x \big( \Box^n \partial_{\alpha_n} T_{\mu_n}   \big) \big( \Box^{n-1} \partial^{\alpha_n} \partial^{\mu_n} \partial^{\sigma_0} T_{\sigma_0} \big) = 2\int d^D x \big(\Box^n T_{\mu_n} \big) \big( \Box^n \partial^{\mu_n} \partial^{\sigma_0} T_{\sigma_0} \big)\,.
\end{equation}
The third term is obtained by integration by parts for three times,
\begin{equation} \label{eq:ThirdTerm}
\begin{aligned}
&\quad \int d^D x \big(\Box^{n-1} \partial_{\alpha_n} \partial_{\mu_n} \partial_{\mu_0} T^{\mu_0} \big) \big( \Box^{n-1}  \partial^{\alpha_n} \partial^{\mu_n} \partial^{\sigma_0} T_{\sigma_0} \big) \\
&= \int d^D x \big( \Box^{n-1} \partial_{\mu_0} T^{\mu_0} \big) \Box^{n-1} \big( \Box^2 \partial^{\sigma_0}T_{\sigma_0} \big) \\
& = - \int d^D x \big( \Box^{n-1} T^{\mu_0} \big) \big( \Box^{n+1} \partial_{\mu_0} \partial^{\sigma_0}T_{\sigma_0} \big)\,.
\end{aligned}
\end{equation}
Therefore the action of the first term is
\begin{equation} \label{eq:GeneralProofPeriodicGaugeField1}
\begin{aligned}
 &\quad-\frac{1}{2} \int d^D x \partial_{\alpha_n} T_{\mu_n} \partial^{\alpha_n} T^{\mu_n}\\
& = \frac{1}{2^{2n+1}} \int d^D x \Big[  \big(\Box^n T_{\mu_n} \big) \big(\Box^{n+1} T^{\mu_n} \big) -2\big(\Box^n T_{\mu_n} \big) \big( \Box^n \partial^{\mu_n} \partial^{\beta_0} T_{\beta_0} \big) +  \big( \Box^{n-1} T^{\mu_0} \big) \big( \Box^{n+1} \partial_{\mu_0} \partial^{\sigma_0}T_{\sigma_0} \big)\Big]\,.\\
\end{aligned}
\end{equation}
Next we will evaluate the second term of the action. 
\begin{equation}
\begin{aligned}
&\quad -\frac{1}{2  } \int d^D x \big( \partial_{\mu} T_{n\, \nu} \partial^{\nu} T_{n}^{\mu} \big) \\
&= -\frac{1}{2^{2n +1}} \int d^D x  \big[ \big( \Box^n  \partial_{\alpha_n}T_{\mu_n} \big) \big( \Box^n \partial^{\mu_n} T^{\alpha_n}\big) -2 \big( \Box^n \partial_{\mu_n} T_{\alpha_n}   \big) \big( \Box^{n-1} \partial^{\alpha_n} \partial^{\mu_n} \partial^{\sigma_0} T_{\sigma_0} \big) \\
& \quad +\big(\Box^{n-1} \partial_{\alpha_n} \partial_{\mu_n} \partial_{\mu_0} T^{\mu_0} \big) \big( \Box^{n-1}  \partial^{\alpha_n} \partial^{\mu_n} \partial^{\sigma_0} T_{\sigma_0} \big)\big]  \,.\\
\end{aligned}
\end{equation}
The first term gives
\begin{equation}
\int d^D x  \big( \Box^n  \partial_{\alpha_n}T_{\mu_n} \big) \big( \Box^n \partial^{\mu_n} T^{\alpha_n}\big) = - \int d^D x \big(\Box^n T_{\mu_n} \big) \big( \Box^n \partial^{\mu_n} \partial_{\alpha_n} T^{\alpha_n} \big)  \,.
\end{equation}
The second term gives
\begin{equation} 
-2\int d^D x  \big( \Box^n \partial_{\alpha_n} T_{\mu_n}   \big) \big( \Box^{n-1} \partial^{\alpha_n} \partial^{\mu_n} \partial^{\sigma_0} T_{\sigma_0} \big)= 2 \int d^D x  \big( \Box^n T_{\mu_n} \big) \big( \Box^n  \partial^{\mu_n} \partial^{\sigma_0} T_{\partial_{\sigma_0}} \big)\,.
\end{equation}
The third term is same as (\ref{eq:ThirdTerm}) in the first-term action case. Hence the second term of the action is
\begin{equation} \label{eq:GeneralProofPeriodicGaugeField2}
\begin{aligned}
&\quad -\frac{1}{2  } \int d^D x \big( \partial_{\mu} T_{n\, \nu} \partial^{\mu} T_{n}^{\nu} \big) \\
&= \frac{1}{2^{2n+1}} \int d^D x \Big[ \big(\Box^n T_{\mu_n} \big) \big( \Box^n \partial^{\mu_n} \partial_{\alpha_n} T^{\alpha_n} \big) -2    \big( \Box^n T_{\mu_n} \big) \big( \Box^n  \partial^{\mu_n} \partial^{\sigma_0} T_{\partial_{\sigma_0}} \big) +   \\  
& \quad\quad\quad\quad\quad\quad\quad\quad\quad\quad\quad\quad\big( \Box^{n-1} T^{\mu_0} \big) \big( \Box^{n+1} \partial_{\mu_0} \partial^{\sigma_0}T_{\sigma_0} \big)\Big]\,.\\
\end{aligned}
\end{equation}
Finally, the full action for gauge field $T_{n}^{\mu}$ is the first term of the action (\ref{eq:GeneralProofPeriodicGaugeField1}) minus the second term of the action (\ref{eq:GeneralProofPeriodicGaugeField2}). Comparing (\ref{eq:GeneralProofPeriodicGaugeField1}) and (\ref{eq:GeneralProofPeriodicGaugeField2}), we find that the last two terms of both expressions are identical and hence cancel. Therefore we have
\begin{equation}
\begin{aligned}
S & = \frac{1}{2^{2n+1}} \int d^D x \Big[ \big(\Box^n T_{\mu_n} \big) \big(\Box^{n+1} T^{\mu_n} \big) - \big(\Box^n T_{\mu_n} \big) \big( \Box^n \partial^{\mu_n}  \partial_{\alpha_n} T^{\alpha_n} \big)\Big] \\
& = \frac{1}{2^{2n+1}} \int d^D x \, \big( \Box^n T^{\mu_n} \big) \big( \Box \eta_{\mu_n \alpha_n} - \partial_{\mu_n} \partial_{\alpha_n} \big) \big( \Box^n T^{\alpha_n} \big) \,.\\
\end{aligned}
\end{equation}
Therefore we obtain the action 
\begin{equation}  \label{eq:result}
\begin{aligned}
S = -\frac{1}{4} \int d^D x G_{n\,\mu\nu} G^{\mu\nu}_n &= \frac{1}{2^{2n+1}} \int d^D x \, \big( \Box^n T^{\mu} \big) \big( \Box \eta_{\mu \nu} - \partial_{\mu} \partial_{\nu} \big) \big( \Box^n T^{\nu} \big)  \\
& = \frac{1}{4^n}\int d^D x \big( \Box^n T^{\mu} \big) \hat{R}_{\mu\nu}  \big( \Box^n T^{\nu} \big)\,.
\end{aligned}
\end{equation}
Then this completes the proof of the most generalized $n$ case, and it is clear that when $n=0$, this regenerates the standard gauge field action that we are familiar with. Under the generalized rotor model of gauge fields, the vertex tensor is coupled to the $\Box^n T^{\mu}$ and $\Box^n T^{\nu}$ fields, contrast to $T^{\mu} $ and $T^{\nu}$ in the standard case. Therefore, under the $n^{\mathrm{th}}$ rotation this amounts to the transformation of fields as
\begin{equation}
 T^{\mu} \rightarrow \Box^n  T^{\mu} \,. 
\end{equation}
And since
\begin{equation}
\Box^n G_{\mu\nu} = \partial_{\mu} \Box^n T_{\nu} - \partial_{\nu} \Box^n T_{\mu} \,,
\end{equation}
it follows that the gauge field strength transforms as,
\begin{equation} \label{eq:transform}
 G_{\mu\nu} \rightarrow \Box^n  G^{\mu\nu} \,. 
\end{equation}
By the transformation in (\ref{eq:transform}), we expect to write down the generalized action for the abelian gauge field under rotor model as
\begin{equation}
S= -a \int d^{D}x \,\Box^n G_{\mu \nu} \Box^n G^{\mu \nu}
\end{equation}
for some constant $a$ that is to be determined. We can check that in fact this form is correct by verifying
\begin{equation} \label{eq:proof}
\begin{aligned}
S&=- a \int d^{D}x \,\Box^n G_{\mu \nu} \Box^n G^{\mu \nu} \\
&=-2a\int d^{D}x \Big( (\partial_{\mu}\Box^n T_{\nu})( \partial^{\mu}\Box^n T^{\nu}) - (\partial_{\mu}\Box^n T_{\nu})( \partial^{\nu}\Box^n T^{\mu}) \Big) \\
&=2a \Bigg(-(\partial_{\mu}\Box^n T_{\nu})\Box^n T^{\nu}\bigg\vert_{-\infty}^{+\infty}+\int d^{D}x\, (\Box^n T^{\nu}) \partial_{\mu}\partial^{\mu}\Box^n T_{\nu } \\
&\quad + (\partial_{\mu}\Box^n T_{\nu})\Box^n T^\mu \bigg\vert_{-\infty}^{+\infty} -\int d^{D}x\, (\Box^n T^\mu) \partial^\nu \partial_{\mu}\Box^n T_{\nu} \Bigg) \\
&= 4a\int d^{D} x \big( \Box^n T^{\mu} \big) \hat{R}_{\mu\nu}  \big( \Box^n T^{\nu} \big) \,,
\end{aligned}
\end{equation}
where in the third step we carry out integration by parts and the boundary terms vanish. Comparing terms in (\ref{eq:result}) and the last line of (\ref{eq:proof}), we equate $4a = \frac{1}{4^n}$ thus this gives $a=
\frac{1}{4^{n+1}}$. Hence, the generalized abelian gauge field theory under rotor model is
\begin{equation}
S = -\frac{1}{4} \int d^{D} x \,G_{n\,\mu\nu} G^{\mu\nu}_n =- \frac{1}{4^{n+1}} \int d^{D}x \,\Box^n G_{\mu \nu} \Box^n G^{\mu \nu} \,,
\end{equation}
where
\begin{equation}
\,G_{n\,\mu\nu} \equiv \frac{1}{2^n} \Box^n G_{\mu\nu} \,.
\end{equation}
Again we can see that when $n=0$, this returns back to the conventional abelian gauge field theory.

From the standard gauge field theory, it is renormalizable with the gauge field dimension as 1 (in terms of the unit of mass). We can work out the dimension $D$ such that the gauge field dimension is retained to be unity. Since the action is dimensionless, it follows that by using (\ref{eq:result}),
\begin{equation}
-D + 2(2n) + 2[T^{\mu}] +2= 0 \,.
\end{equation}
If we pick $[T^{\mu}]=1$, then we demand $D=4n+4$. It can be easily checked that when $n=0$, $D=4$ which is the standard case.

\section{Equation of motion and Noether's current of the generalized abelian gauge field theory}
For the $n=0$ case, we know that the equation of motion is the Maxwell equation,
\begin{equation} \label{eq:eom1}
\partial_{\mu} G^{\mu\nu} = 0 \,,
\end{equation}
which can be obtained by varying the first equation of (\ref{eq:standard}). Alternatively we can write it as
\begin{equation} \label{eq:eom2}
\hat{R}_{\mu\nu} T^{\nu} = 0 \,.
\end{equation}
which can be obtained by varying the second equation of (\ref{eq:standard}). To derive the equation of motion of the generalized abelian gauge field action, we vary the second equation in (\ref{eq:general}),
\begin{equation} \label{eq:eq1}
\begin{aligned}
\delta S &= \frac{1}{4^n} \int d^{D} x \Big( (\delta \Box^n T^{\mu})\hat{R}_{\mu\nu} \Box^n T^{\nu} + \Box^n T^{\mu}\delta \hat{R}_{\mu\nu}  \Box^n T^\nu+ \Box^n T^{\mu}\hat{R}_{\mu\nu} \delta \Box^n T^\nu \Big) \\
&= \frac{1}{4^n} \int d^{D} x \Big((\delta \Box^n T^{\mu})\hat{R}_{\mu\nu} \Box^n T^{\nu} + \frac{1}{2}\Box^n T^{\mu} \Box \eta_{\mu\nu}\delta\Box^n T^{\nu} -\frac{1}{2}\Box^n T^{\mu}\partial_{\mu}\partial_{\nu}\delta\Box^n T^{\nu}\Big) \\
&= \frac{1}{4^n} \int d^{D} x \Big((\delta \Box^n T^{\mu})\hat{R}_{\mu\nu} \Box^n T^{\nu} + \frac{1}{2}\Box^n T^{\nu} \Box \eta_{\nu\mu}\delta\Box^n T^{\mu} -\frac{1}{2}\Box^n T^{\nu}\partial_{\nu}\partial_{\mu}\delta\Box^n T^{\mu}\Big) \,,
\end{aligned}
\end{equation}
where in the first line $\delta \hat{R}_{\mu\nu} =0$ as $ \hat{R}_{\mu\nu}$ is independent of $\Box^{n}T^{\mu}$ fields. Since in the last line the last two terms in (\ref{eq:eq1}) involve second order derivatives, we cannot simply use integration by parts directly. Instead we need to derive an integration rule for the second derivative case. First we would like to compute the integral 
\begin{equation} \label{eq:eq0}
\int d^{D} x\, \Box^n T^{\nu} \Box \eta_{\nu\mu}\delta\Box^n T^{\mu} \,.
\end{equation}
Define a second rank tensor $W^{\nu\mu}$ such that
\begin{equation}
W^{\nu \mu} = a^{\nu} b^{\mu}
\end{equation}
for some arbitrary $a^{\nu}, b^{\mu}$ rank-1 tensors. Consider 
\begin{equation} \label{eq:eq00}
\partial^\rho \partial^{\sigma}W^{\nu \mu} = b^{\mu} ( \partial^{\rho}\partial^{\sigma}a^\nu  ) + ( \partial^\sigma a^\nu )( \partial^\rho b^\mu) +( \partial^\rho a^\nu) (\partial^\sigma b^\mu) + a^{\nu} ( \partial^{\rho}\partial^{\sigma} b^{\mu}   ) \,.
\end{equation} 
Rearranging and integrating both sides, it follows that,
\begin{equation}
\begin{aligned}
\int d^{D} x a^{\nu} ( \partial^{\rho}\partial^{\sigma} b^{\mu}) &= \int d^{D} x \,\partial^\rho \partial^{\sigma}W^{\nu \mu} - \int d^{D} x \, b^{\mu} ( \partial^{\rho}\partial^{\sigma}a^\nu  ) \\
&\quad- \int d^{D} x\, ( \partial^\sigma a^\nu )( \partial^\rho b^\mu)- \int d^{D} x \, ( \partial^\rho a^\nu) (\partial^\sigma b^\mu) \,.
\end{aligned}
\end{equation}
Substituting $a^\nu =\Box^n T^{\nu}$ and $b^{\mu} = \delta \Box^n T^\mu$, then
\begin{equation} \label{eq:eq2}
\begin{aligned}
\int d^{D} x \Box^n T^{\nu} ( \partial^{\rho}\partial^{\sigma} \delta \Box^n T^\mu) &= \int d^{D} x \,\partial^\rho \partial^{\sigma} (\Box^n T^{\nu }\delta \Box^n T^\mu) - \int d^{D} x  \delta \Box^n T^\mu ( \partial^{\rho}\partial^{\sigma} \Box^nT^\nu  ) \\
&\quad- \int d^{D} x\, ( \partial^\sigma\Box^n T^\nu )( \partial^\rho \delta \Box^n T^\mu)- \int d^{D} x \, ( \partial^\rho \Box^n T^\nu) (\partial^\sigma \delta \Box^n T^\mu) \,.
\end{aligned}
\end{equation}
The first term in (\ref{eq:eq2}) is the boundary term that vanishes. Using integration by parts for the third and the forth term, then we have
\begin{equation}
\begin{aligned}
\int d^{D} x \Box^n T^{\nu} ( \partial^{\rho}\partial^{\sigma} \delta \Box^n T^\mu) &= - \int d^{D} x  \delta \Box^n T^\mu ( \partial^{\rho}\partial^{\sigma} \Box^n T^\nu  )\\
& \quad- \bigg( (\partial^\sigma \Box^n T^\nu )\delta \Box^n T^\mu \bigg\vert^{+\infty}_{-\infty}-\int d^{D}x\,(\delta \Box^n T^\mu) \partial^{\rho}\partial^{\sigma} \Box^nT^{\nu}\bigg) \\
&\quad - \bigg( (\partial^\rho \Box^n T^\nu )\delta \Box^n T^\mu \bigg\vert^{+\infty}_{-\infty}-\int d^{D}x\,(\delta \Box^n T^\mu ) \partial^{\sigma}\partial^{\rho} \Box^n T^{\nu}\bigg) \\
&=\int d^{D} x  (\delta \Box^n T^\mu ) \partial^{\rho}\partial^{\sigma} \Box^n T^\nu \,.
\end{aligned}
\end{equation}
In the second and the third line, the boundary terms are zero thus they vanish. Now multiply both sides by the metric tensors $\eta_{\rho\sigma}\eta_{\mu\nu}$, and as $\eta_{\rho\sigma} \partial^{\rho}\partial^{\sigma} =\Box$, thus the integral in (\ref{eq:eq0}) is evaluated to be
\begin{equation} \label{eq:result1}
\int d^{D} x\, \Box^n T^{\nu} \Box \eta_{\nu\mu}\delta\Box^n T^{\mu} = \int d^{D} x  (\delta \Box^n T^\mu )  \Box \eta_{\mu\nu} \Box^n T^\nu \,.
\end{equation}
Similarly, modifying equation (\ref{eq:eq00}) to $\partial_{\nu}\partial_{\mu} W^{\nu\mu}$, and using the same technique, we will obtain, for the third term in the last line of equation (\ref{eq:eq1}) as
\begin{equation} \label{eq:result2}
\int d^{D} x \Box^n T^{\nu}\partial_{\nu}\partial_{\mu}\delta\Box^n T^{\mu} = \int d^{D} x (\delta\Box^n T^{\mu})\partial_{\nu}\partial_{\mu}\Box^n T^{\nu} \,.
\end{equation}
Therefore it follows from equations (\ref{eq:eq1}) and (\ref{eq:result1}) and (\ref{eq:result2}), we have
\begin{equation}
\begin{aligned}
\delta S &= \frac{1}{4^n} \int d^{D} x \Big((\delta \Box^n T^{\mu})\hat{R}_{\mu\nu} \Box^n T^{\nu} + \frac{1}{2} (\delta \Box^n T^\mu )  \Box \eta_{\mu\nu} \Box^n T^\nu - \frac{1}{2} (\delta\Box^n T^{\nu})\partial_{\nu}\partial_{\mu}\Box^n T^{\mu} \bigg) \\
&=\frac{1}{4^n} \int d^{D} x \Big((\delta \Box^n T^{\mu})\hat{R}_{\mu\nu} \Box^n T^{\nu} + (\delta \Box^n T^{\mu})\frac{1}{2}(\Box\eta_{\mu\nu}-\partial_{\mu}\partial_{\nu})\Box^n T^{\nu}) \\
&=\frac{1}{4^n}\cdot 2  \int d^{D} x (\delta \Box^n T^{\mu})\hat{R}_{\mu\nu} \Box^n T^{\nu} \,.
\end{aligned}
\end{equation}
The minimization of action is achieved by
\begin{equation}
\frac{\delta S}{\delta (\Box^n T^\mu)} = 0 \,.
\end{equation}
Therefore, the equation of motion of the generalized gauge field is
\begin{equation} \label{eq:EOM}
\hat{R}_{\mu\nu} \Box^n T^{\nu} = 0 \,.
\end{equation}
Since partial derivative commutes, we can write
\begin{equation}
\Box^n  \hat{R}_{\mu\nu} T^{\nu} =0 \,.
\end{equation}
Alternatively we can write
\begin{equation}
  \partial_{\mu} ( \Box^n G^{\mu\nu}) =\Box^n \partial_{\mu} G^{\mu\nu}  = 0 \,.
\end{equation}
Again, when $n=0$ this gives us back the original Maxwell equation. Alternatively, the equation of motion can be obtained by the Euler Lagrange equation. We have the action with field variables of $\Box^n T_{\nu}$ and $\partial_{\mu} \Box^n T_{\nu}$,
\begin{equation}
    S= \int d^D x \mathcal{L}(\Box^n T_{\nu}, \partial_{\mu} \Box^n T_{\nu} ) \,.
\end{equation}
The variation of action gives
\begin{equation}
 \begin{aligned}
 \delta S &= \int d^D x \bigg(\frac{\partial \mathcal{L}}{\partial \Box^n T_{\nu}} \delta \Box^n T_{\nu} + \frac{\partial \mathcal{L}}{\partial ( \partial_{\mu} \Box^n T_{\nu} )} \delta (\partial_{\mu} \Box^n T_{\nu}) \bigg) \\
 &= \int d^D x \bigg(\frac{\partial \mathcal{L}}{\partial \Box^n T_{\nu}} \delta \Box^n T_{\nu} + \frac{\partial \mathcal{L}}{\partial ( \partial_{\mu} \Box^n T_{\nu} )}  \partial_{\mu} \delta \Box^n T_{\nu} \bigg) \\
 &=\int d^D x \bigg(\frac{\partial \mathcal{L}}{\partial \Box^n T_{\nu}}\delta \Box^n T_{\nu} \bigg) + \frac{\partial \mathcal{L}}{\partial (  \partial_{\mu} \Box^n T_{\nu}  )}  \delta \Box^n T_{\nu} \bigg\vert_{-\infty}^{+\infty} - \int d^D x (\delta \Box^n T_{\nu}) \partial_{\mu} \frac{\partial \mathcal{L}}{\partial ( \partial_{\mu} \Box^n T_{\nu} )} \\
 &= \int d^D x  \delta \Box^n T_{\nu}  \bigg(\frac{\partial \mathcal{L}}{\partial \Box^n T_{\nu}} - \partial_{\mu} \frac{\partial \mathcal{L}}{\partial ( \partial_{\mu} \Box^n T_{\nu} )} \bigg) \,,
 \end{aligned}
\end{equation}
where in the third line the boundary term vanishes. The minimization of the action takes place when $\delta S =0$. Then we obtain the Euler Lagrange equation as
\begin{equation} \label{eq:EL}
\frac{\partial \mathcal{L}}{\partial \Box^n T_{\nu}} =   \partial_{\mu} \frac{\partial \mathcal{L}}{\partial ( \partial_{\mu} \Box^n T_{\nu} )} \,.
\end{equation}
It can be checked that using (\ref{eq:EL}) can give us back the equation of motion in (\ref{eq:EOM}).

Finally we compute the Noether's current of our theory. Using the Euler Lagrange equation in (\ref{eq:EL}), we have
\begin{equation}
\begin{aligned}
     0  &= \delta \mathcal{L} \\
        &= \bigg(\partial_{\mu} \frac{\partial \mathcal{L}}{\partial ( \partial_{\mu} \Box^n T_{\nu} )} \bigg) \delta \Box^n T_{\nu} + \frac{\partial \mathcal{L}}{\partial ( \partial_{\mu} \Box^n T_{\nu} )}  \partial_{\mu} \delta \Box^n T_{\nu} \\
        &= \partial_{\mu} \bigg(  \frac{\partial \mathcal{L}}{\partial ( \partial_{\mu} \Box^n T_{\nu} )} \delta \Box^n T_{\nu} \bigg) \,.
\end{aligned}
\end{equation}
Since the conserved current satisfies $\partial_{\mu} J^{\mu}=0$, therefore we identify the conserved Noether's current as
\begin{equation}
    J^{\mu} =  \frac{\partial \mathcal{L}}{\partial ( \partial_{\mu} \Box^n T_{\nu} )} \delta \Box^n T_{\nu} \,.
\end{equation}
Upon computation,
\begin{equation}
    \frac{\partial \mathcal{L}}{\partial ( \partial_{\mu} \Box^n T_{\nu} )} = -\Box^n G^{\mu\nu} \,,
\end{equation}
then we obtain the Noether's current as
\begin{equation}
     J^{\mu} = \Box^n G^{\nu\mu} \delta \Box^n T_{\nu}\,.
\end{equation}
The gauge transformation is given by
\begin{equation}
    T_{\nu}^\prime =  T_{\nu} + \partial_{\nu} \theta (x) \,,
\end{equation}
where $\theta(x)$ is some scalar function. It follows that
\begin{equation}
    \Box^n T_{\nu}^\prime = \Box^n T_{\nu} + \partial_{\nu}\Box^n  \theta (x)\,.
\end{equation}
Hence we have the infinitesimal change of the rotor $\Box^n T_{\nu}$ field as
\begin{equation}
    \delta \Box^n T_{\nu}=  \Box^n T_{\nu}^\prime - \Box^n T_{\nu}  =\partial_{\nu}\Box^n  \theta (x) \,.
\end{equation}
Therefore the Noether's conserved current is
\begin{equation}
     J^{\mu} = \Box^n G^{\nu\mu} \partial_{\nu}\Box^n  \theta \,.
\end{equation}
The associated Noether's charge $Q$ is
\begin{equation}
 Q = \int d^{D-1} x J^{0} = \int d^{D-1} x \Box^n G^{\nu 0} \partial_{\nu}\Box^n  \theta \,.
\end{equation}

The extended Lorentz gauge condition for the rotor model is
\begin{equation}
    \partial^{\mu} \Box^n T_{\mu} = 0 \,,
\end{equation}
which demands
\begin{equation}
    \Box^{n+1}\theta = 0 \,.
\end{equation}
Again we can see that when $n=0$, it returns to the original standard gauge field case.

\section{Conclusion}
We have extended the original standard gauge field theory to its generalized version under the rotor model of fields of $n^{\mathrm{th}}$ order and established the generalized abelian gauge field theorem. The Feynman vertex is interpreted as the projection tensor, which couples to the oscillation fields $\Box^n T^{\mu}$ and $\Box^n T^{\nu}$. We demonstrated how the action of rotations given by the second order rotational operators can increase the order of derivatives in the Lagrangian. We show that generally, under such rotation the gauge field and gauge field strength are transformed by $T^{\mu} \rightarrow \Box^n T^{\mu} $ and $G^{\mu \nu} \rightarrow \Box^n G^{\mu\nu} $ respectively. The dimension of the gauge field in such generalized theory is $4n+4$ in order to be renormalizable with unity field dimension. We also derive the equation of motion and Noether's conserved current of the generalized abelian gauge field theory. The $n=0$ trivial case returns to the standard gauge field theory itself. In future, we will investigate the possibility of extending our rotor model formalism to non-abelian gauge field theory. In addition, the formalism of generalized abelian gauge field model under rotor model in curved spacetime can be studied. However, the work will be far much harder because covariant derivatives, unlike partial derivatives, do not commute and swapping covariant derivatives will amount to a Riemannian curvature tensor. The theory in curved spacetime will be far much more complicated.

\end{document}